\documentclass[sigconf]{acmart}
\AtBeginDocument{%
  }

\usepackage{multirow,tabularx} 
\usepackage{adjustbox} 
\usepackage{booktabs}
\usepackage{color, colortbl} 
\usepackage{array, booktabs, makecell}
\usepackage{comment} 
\usepackage{color,soul}
\usepackage{graphicx}
\usepackage{caption}
\usepackage{subcaption}
\usepackage{nth}
\usepackage{url}
\usepackage{textcomp}
\usepackage[normalem]{ulem}
\usepackage{pgfplots}
\usepackage{tikz}
\usetikzlibrary{patterns}
\usepackage[figuresright]{rotating}
\usepackage[most]{tcolorbox}   
\usepackage{textcomp}  
\usepackage{mathtools}
\usepackage{amsmath}
\usepackage{tabularx}
\usepackage{pgfplots}
\usepackage{pgfplotstable}
\usepackage{sfmath}
\usepackage{array, booktabs, makecell}
\usepackage{color}
\usepackage{csquotes}
\usepackage{url}
\usepackage{comment}
\usepackage{tikz}
\usepackage{listings}
\usepackage{breqn}
\usepackage{booktabs} 
\usepackage{multirow}
\usetikzlibrary{fit} 
\usetikzlibrary{positioning}
\usetikzlibrary{arrows}
\usetikzlibrary{shapes.multipart}
\usepackage{float}
\usepackage{enumitem}
\usepackage[small,compact]{titlesec}
\usepackage{mathtools}
\usepackage{pgfplots}
\usepgfplotslibrary{fillbetween}
\pgfplotsset{compat=1.11,width=10cm}
\usepgfplotslibrary{statistics}

  {\list{}{\leftmargin=0.1in\rightmargin=0.1in}\item[]}%
  {\endlist}

\usepackage{pgfplots, pgfplotstable}

\usepackage{bchart}
\usepackage{balance}
\usepackage{pgf-pie}  
\usepackage{tabularx}
\usepackage{pgfplots}
\usepackage{multicol}
\usepackage{caption}

\usetikzlibrary{decorations.pathmorphing}
\tikzset{snake it/.style={decorate, decoration=snake}}
\usepackage{xcolor}
\definecolor{MyOrange}{RGB}{237,125,49}
\definecolor{MyBlue}{RGB}{91,155,213}

\definecolor{light-gray}{rgb}{0.95,0.95,0.95}

\newtcolorbox{box1}[1][]
{
  enhanced,
  sharpish corners,
  colback=light-gray,
  colframe=light-gray,
  top=0.05cm,
  bottom=0.05cm,
  left=0.05cm,
  right=0.05cm,
  enlarge top by=0.2cm,
  enlarge bottom by=0.2cm,
  breakable
}

\AtBeginDocument{%
  }

\usepackage[flushleft]{threeparttable}
\usepackage{tikz}

\usepackage{booktabs}
\usepackage{ragged2e}
\usepackage[table]{xcolor}
\usepackage{pifont} 
\newcommand{\cmark}{\ding{51}}
\usetikzlibrary{patterns,patterns.meta}

\newcolumntype{Y}{>{\RaggedRight\arraybackslash}X}

\newtcolorbox{box2}[1][]
{
  title=#1, 
  enhanced,
  colback=blue!20,
  frame hidden,
  colbacktitle=gray, 
  boxed title style={colframe=gray},
  top=0.05cm,
  bottom=0.05cm,
  left=0.15cm,
  right=0.15cm,
  enlarge top by=0.5cm,
  enlarge bottom by=0.3cm,
  attach boxed title to top left={xshift=3mm,
  yshift=-3mm,yshifttext=-1mm},
  borderline west={4pt}{0pt}{blue!50!black},
  breakable
}

\copyrightyear{2026}
\acmYear{2026}
\setcopyright{cc}
\setcctype{by}
\acmConference[W4A '26]{Web4All 2026 23rd International Web for All Conference}{April 13--14, 2026}{Dubai, United Arab Emirates}
\acmBooktitle{Web4All 2026 23rd International Web for All Conference (W4A '26), April 13--14, 2026, Dubai, United Arab Emirates}
\acmDOI{10.1145/3800424.3800452}
\acmISBN{979-8-4007-2372-8/2026/04}

\begin{document}

\title{Large Language Models for Web Accessibility: A Systematic Literature Review}

\newcommand\Mark[1]{\textsuperscript#1}

\author{Wajdi Aljedaani}
\email{waljedaani@sdaia.gov.sa}
\orcid{0000-0002-6700-719X}
\affiliation{%
 \institution{Saudi Data \& AI Authority}
 \city{Riyadh}
\country{KSA}
}

\author{Rubel Hassan Mollik}
\email{RubelHassanMollik@my.unt.edu}
\orcid{0009-0004-9714-1877}
\affiliation{
 \institution{University of North Texas}
 \city{Denton}\state{Texas}
\country{USA}
\postcode{}
}

\begin{abstract}

Web accessibility aims to ensure that web content and services are usable by people with diverse abilities. In recent years, Large Language Models (LLMs) have been increasingly explored to support accessibility-related tasks on the web, such as content generation, issue detection, and remediation. However, little is known about the characteristics of these approaches, the accessibility issues they target, the standards they follow, and how they are evaluated. In this paper, we present a systematic literature review of 38 peer-reviewed studies that investigate the use of LLMs in web accessibility contexts. We begin by performing a comprehensive search of scientific publications to identify relevant studies. We then conduct a comparative analysis to examine the accessibility tasks addressed, the LLM models and prompting strategies employed, the system architectures adopted, the accessibility issues and guidelines considered, and the evaluation methods used across studies. Our findings show that most studies apply LLMs to text-centric and structurally explicit accessibility tasks, with WCAG serving as the primary reference framework and limited consideration of cognitive accessibility guidelines (COGA). The reviewed approaches predominantly rely on general-purpose LLMs and prompt-based interactions, while evaluation practices vary widely and often lack direct involvement of users with disabilities. We envision this review as a consolidated reference for researchers and practitioners seeking to understand the current landscape of LLM-supported web accessibility, and as a foundation to guide future research and tool development in this area.

\end{abstract}

\ccsdesc[500]{Human-centered computing~ Accessibility}

\keywords{Accessibility, Web Accessibility, Large Language Models, LLM, Accessibility, WCAG, Artificial Intelligence}



\maketitle

\section{Introduction}
\label{sec:Introduction}
\vspace{-0.3cm}
The rapid advancement of Large Language Models (LLMs) has led to their widespread adoption across web development and content creation workflows~\cite{aljedaani2024does,zhou2025examining}. These models are now routinely used to generate web code, assist with interface design, produce textual and visual descriptions, and support interactive web-based systems~\cite{fathallah2025accessguru,lopez2025turning,mehendale2024dexassist}. As LLMs increasingly shape how web content is authored and delivered, their influence extends beyond functionality and aesthetics to fundamental quality attributes of the web, including accessibility.

Web accessibility concerns the extent to which web content and services can be accessed and used by people with diverse abilities~\cite{lewthwaite2014web}. Although accessibility standards and best practices have matured over time, accessibility remains inconsistently implemented across the web~\cite{webaim2020webaim}. Many websites continue to present barriers for users with disabilities~\cite{hewitt2021internet,aljedaani2025challenges}, particularly when accessibility considerations are deferred or treated as optional during development. Addressing accessibility often requires specialized expertise, careful evaluation, and sustained maintenance, which can be perceived as time-consuming or difficult to scale within fast-paced development environments~\cite{coverdale2024digital}. 

In response to these challenges, researchers have begun investigating whether LLMs can be leveraged to support accessibility-related tasks on the web. Existing studies explore a range of applications, including generating alternative text for images, detecting accessibility violations, assisting with the creation of accessible markup, explaining accessibility guidelines, and supporting users through conversational interfaces~\cite{fathallah2025accessguru,mowar2025codea11y,lopez2025turning,mehendale2024dexassist}. These works demonstrate growing interest in using LLMs to reduce the effort associated with accessibility work. At the same time, reported findings vary widely, with some studies highlighting promising results~\cite{fathallah2025accessguru} and others identifying substantial limitations related to accuracy, consistency, and reliability~\cite{aljedaani2024does}. 

Despite a growing body of work, research on LLMs and web accessibility remains fragmented. Existing studies often focus on narrow problem settings, such as individual accessibility tasks~\cite{moon2024sagol,mo2025tablenarrator, li2025videoa11y}, specific web technologies~\cite{fathallah2025accessguru,cao2025scenegena11y, doush2024evaluating}, or isolated evaluation techniques~\cite{he2025enhancing, paterno2025llm, andruccioli2025tabular}
Moreover, accessibility is operationalized inconsistently across the literature, with varying standards, metrics, and validation practices~\cite{singh2024accessibility, abu2025can, vera2025accessible}. This fragmentation makes it difficult to develop a cohesive understanding of how LLMs, as a broader class of systems, contribute to—or potentially undermine—web accessibility beyond isolated tool-level analyses. At the same time, web accessibility poses unique challenges for LLM-based systems, distinguishing it from other application domains. Accessibility-related outputs must adhere to established standards, address diverse user needs, and avoid generating misleading or superficial guidance~\cite{oswal2024examining, delnevo2024interaction}. Failures such as hallucinated compliance claims or incomplete fixes can have direct and disproportionate consequences for users with disabilities, underscoring the need for a systematic examination of how LLMs are currently evaluated and applied in accessibility-focused web contexts~\cite{gurita2025llm, andruccioli2025leveraging, huang2025survey, droutsas2025web}. 

To address these gaps, 
we conduct a Systematic Literature Review (SLR) of studies that investigate the role of LLMs in web accessibility. 
We analyze 33 primary studies spanning accessibility tasks, web technologies, evaluation methods, and application contexts. 
Our synthesis identifies dominant research trends, commonly reported challenges, and methodological gaps in current LLM-supported accessibility work. This review establishes 
how LLMs are currently used to support web accessibility, where their limitations persist, and 
outlines directions for future research toward more reliable and inclusive web experiences.

\subsection{Goal \& Research Questions}
The goal of this study is to provide researchers, practitioners, and accessibility stakeholders with a comprehensive and structured synthesis of existing research on the use of LLMs in web accessibility. By systematically reviewing and analyzing prior work, this study consolidates fragmented evidence on how LLMs are applied to support accessibility-related web tasks, the standards and disability groups they address, and the methods used to evaluate their effectiveness. The findings aim to inform decision-making, highlight methodological and technical gaps, and guide future efforts toward more reliable, inclusive, and standards-aligned LLM-supported web accessibility solutions. To achieve these objectives, we formulate the following research questions (RQs):

\vspace{-0.6cm}
\begin{box2}
\textbf{RQ$_1$: What are Large Language Models (LLMs) being applied to support web accessibility?}
\end{box2}
\vspace{-0.5cm}

This research question examines the types of accessibility tasks addressed by LLM-based approaches, such as detection, remediation, content generation, and user support, and characterizes how these models are integrated into web accessibility workflows.

\vspace{-0.6cm}
\begin{box2}
\textbf{RQ$_2$: What are LLM-based approaches for web accessibility that incorporate accessibility standards, guidelines, and disability considerations?}
\end{box2}
\vspace{-0.5cm}

This RQ examines the extent to which existing studies explicitly reference and operationalize accessibility standards and guidelines, and the disability groups they consider. Addressing this question allows us to assess alignment with established accessibility frameworks and identify gaps in coverage across user needs.

\vspace{-0.6cm}
\begin{box2}
\textbf{RQ$_3$: What LLM models, prompting strategies, and system architectures are used in web accessibility research?}
\end{box2}
\vspace{-0.5cm}
This research question focuses on the technical design choices made in prior work, including the selection of LLMs, the use of prompting techniques, and the adoption of architectural patterns. By answering this RQ, we provide insight into prevailing design practices and model diversity within the literature.

\vspace{-0.6cm}
\begin{box2}
\textbf{RQ$_4$: What web accessibility issues and WCAG success-criterion violations are most frequently addressed?}
\end{box2}
\vspace{-0.5cm}
This RQ investigates the types of accessibility issues targeted by existing studies and how LLMs are applied to detect, remediate, or generate accessible content for these issues. The findings highlight dominant problem areas and less explored accessibility concerns.

\vspace{-0.6cm}
\begin{box2}
\textbf{RQ$_5$: How are LLM-based web accessibility approaches evaluated, and what evidence is used to assess their effectiveness?}
\end{box2}
\vspace{-0.5cm}
This research question examines the evaluation strategies employed across studies, including automated assessments, expert reviews, user studies, and comparative analyses. Addressing this RQ provides insight into the rigor, consistency, and limitations of current evaluation practices.

\subsection{Contributions}

\begin{itemize}[wide, labelwidth=1pt, labelindent=0pt]
    \item A systematic literature review of 33 peer-reviewed studies examining how LLMs are applied to support web accessibility.

    \item A structured characterization of LLM-based accessibility approaches, including the tasks addressed, solution types, models, and interaction strategies used across studies.

    \item An analysis of accessibility standards, disability coverage, and reported issues, highlighting dominant trends and gaps in current research.

    \item A synthesis of evaluation practices and open challenges, offering insights to guide future research toward more reliable and inclusive LLM-supported web accessibility solutions.

    \item A replication package of our survey for extension purposes \footnote{Will be shared upon acceptance}.
\end{itemize}

The remainder of this paper is organized as follows. Section \ref{sec:RelatedWork} reviews prior systematic literature reviews and related work in web accessibility and human–computer interaction, and identifies gaps addressed by this study. Section \ref{sec:ResearchMethodology} describes the research methodology, detailing the planning, execution, and synthesis phases of the systematic literature review. Section \ref{sec:ResultsAndAnalysis} presents the research findings, structured around the five research questions and summarizing observed patterns in LLM applications, accessibility issues, and evaluation practices. Section \ref{sec:Discussions} discusses the results by synthesizing key themes and challenges identified across the literature. Section \ref{sec:ResearchImplications} outlines the implications of our findings for both research and practice. Section \ref{sec:ThreatsToValidity} discusses threats to validity and the mitigation strategies employed. Finally, Section \ref{sec:Conclusion} concludes the paper by summarizing the main insights and outlining directions for future work.

\section{Related Work}
\label{sec:RelatedWork}

A substantial body of work in the HCI and accessibility communities has employed systematic literature reviews to examine web accessibility practices, evaluation methods, and guideline adoption. Several SLRs focus on web accessibility evaluation tools and methods \cite{ara2024accessibility, campoverde2023accessibility, campoverde2020empirical, ordonez2022model}, analyzing how automated, semi-automated, and manual approaches are used to assess compliance with WCAG and related standards \cite{vigo2013benchmarking}. These studies consistently report limitations of automated tools in capturing contextual, semantic, and user-experience-dependent accessibility issues, emphasizing the need for complementary approaches. Other SLRs investigate accessibility integration within the software development lifecycle \cite{cruz2021accessibility}, highlighting that accessibility is often addressed late in development and inconsistently applied across design \cite{sariouglu2025accessibility}, requirements \cite{teixeira2024understanding}, and testing phases \cite{rajh2024accessibility}. Collectively, these reviews provide valuable insights into accessibility challenges, tooling limitations, and methodological gaps, however, they predate or only tangentially consider the role of modern generative AI systems.

More recent SLR extend this line of inquiry by examining AI-driven approaches for accessibility, including computer vision, speech technologies, and adaptive interfaces \cite{ali2025use}. However, this review typically treat AI as a broad category and does not isolate Large Language Models (LLMs) as a distinct class of systems with unique capabilities and risks. In parallel, SLRs focusing on cognitive accessibility and inclusive design emphasize that COGA-related concerns—such as content complexity, cognitive load, and clarity—remain underrepresented in both research and tooling \cite{borina2022web}. While these works underscore important gaps in accessibility coverage and evaluation rigor, none provide a systematic synthesis of how LLMs are applied to web accessibility tasks, how their outputs align with WCAG~\cite{caldwell2008web} and COGA guidelines~\cite{w3c_coga_usable_2021}, or how LLM-based approaches are evaluated in practice. This gap motivates our systematic literature review, which consolidates evidence across 38 studies to map the landscape of LLM-supported web accessibility research, identify dominant patterns, and highlight underexplored areas within the HCI and accessibility domains.

\begin{figure*}
	\centering
    \includegraphics[width=0.7\textwidth]{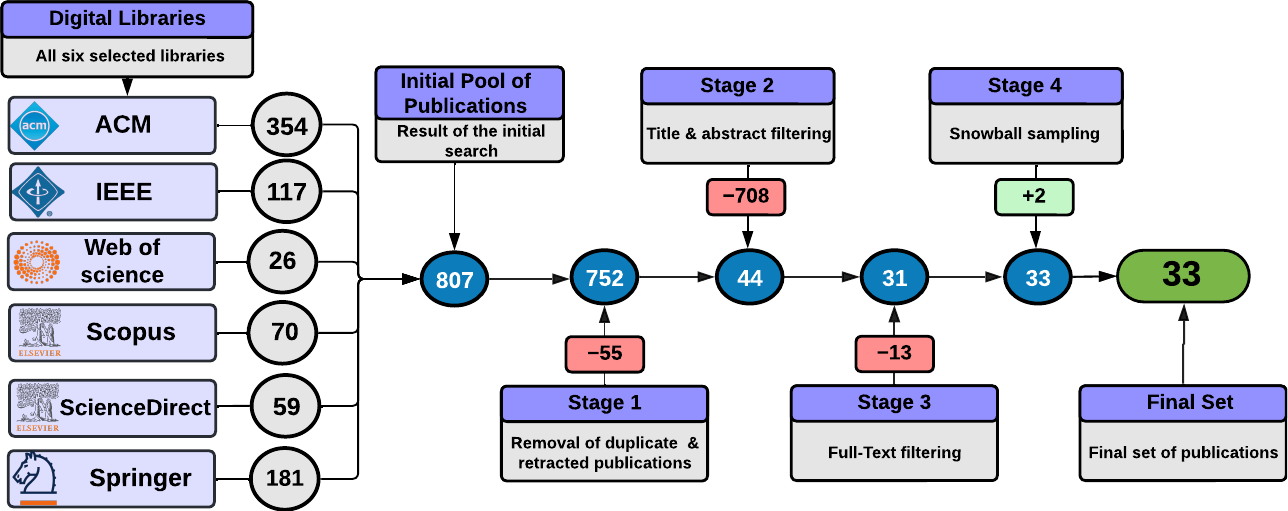}   
    \vspace{-0.3cm}
    \caption{Overview of the volume of publications resulting from our filtering process.}
    \label{fig:SearchProcess}
    \vspace{-0.4cm}
\end{figure*}

\section{Research Methodology}
\label{sec:ResearchMethodology}

As a Systematic Literature Review (SLR), this study systematically reviews and synthesizes the existing scientific literature to provide a structured, comprehensive understanding of how Large Language Models (LLMs)~\cite{naveed2025comprehensive} are applied in web accessibility~\cite{brophy2007web}. SLRs~\cite{kitchenham2009systematic} are widely used in software engineering and human–computer interaction research to consolidate fragmented evidence, identify research trends, and highlight gaps in emerging domains. In this work, we aim to analyze prior studies on the use of LLMs to support, evaluate, or enhance web accessibility and derive insights to inform future research and practice.

While prior reviews have explored LLMs in broader software engineering contexts or examined accessibility independently, this study focuses specifically on the intersection of LLMs and web accessibility. Accordingly, this section describes the methodology we adopted to identify, select, and analyze relevant publications. Our review process consists of three main phases: (1) planning, (2) execution, and (3) synthesis. Each phase is described in detail in the following subsections.

\subsection{Planning}
The planning phase defines the scope of the review and establishes a systematic strategy for identifying relevant studies. This phase includes selecting digital libraries, defining inclusion and exclusion criteria, and constructing search keywords.

\textbf{Digital Libraries.}
To locate publications for our study, we searched six digital libraries. These libraries, listed in Table 1, either contain or index peer-reviewed publications from computer science, software engineering, and human–computer interaction venues, and have been commonly utilized in prior systematic literature reviews (e.g., [32]).

\begin{tcolorbox}[colback=black!5!white,colframe=black!50!black,top=2mm, bottom=2mm, left=2mm, right=2mm]
\texttt{\emph{("web accessib*" OR "website* accessib*" OR "web content accessib*" OR "digital accessib*" OR "online accessib*" OR "internet accessib*" OR "accessible web" OR "accessible website*" OR "accessible web page*" OR "accessible web design" OR "web content accessibility guidelines" OR "wcag" OR "wai-aria" OR "a11y" )
AND
("large language model" OR "large language models" OR "llm" OR "llms" OR "llm-based" OR "foundation model" OR "foundation models" OR "chatgpt" OR "gpt" OR  "generative ai" OR "generative artificial intelligence" OR "vision-language model" OR "multimodal model")}}
\end{tcolorbox}

\textbf{Search Keywords.}
To identify an appropriate set of search keywords, we conducted a pilot search across two widely used digital libraries: IEEE Xplore and the ACM Digital Library. This process was used to identify commonly used terms and synonyms across studies on Large Language Models and web accessibility. The search query was refined through multiple iterations to improve relevance and coverage. To reduce false positives, we applied the finalized search query only to publication titles and abstracts, rather than the full text. The finalized search string used across all selected digital libraries is presented below.

\begin{table}
\centering
\caption{The digital libraries queried in our study. }
\vspace{-0.3cm}
\small
\label{tab:digital_libraries}
\begin{tabular}{|l|l|}\hline
\rowcolor{gray!60}\multicolumn{1}{|c|}{\textbf{Digital Library}} & \multicolumn{1}{c|}{\textbf{URL}} \\
ACM Digital Library                            & \url{https://dl.acm.org/}               \\
\rowcolor{gray!30} IEEE Xplore                                     & \url{https://ieeexplore.ieee.org/}      \\
 Web of Science                                 & \url{https://webofknowledge.com/}       \\ 
\rowcolor{gray!30} Scopus                                         & \url{https://www.scopus.com/}           \\
Science Direct                                 & \url{https://www.sciencedirect.com/}    \\

\rowcolor{gray!30}Springer Link                                  & \url{https://link.springer.com/}        \\

\hline
\end{tabular}
\vspace{-0.3cm}
\end{table}

\begin{table}[h!]
\centering
\vspace{-0.2cm}
\caption{Our inclusion and exclusion search criteria.}  
\vspace{-0.3cm}
\resizebox{\columnwidth}{!}{%
\begin{tabular} {|l|l|}\hline

\rowcolor{gray!60}
 \multicolumn{1}{|c|}{\textbf{Inclusion}} & \multicolumn{1}{|c|}{\textbf{Exclusion}} \\ 
\hline 
Peer-reviewed publication	& Grey literature (e.g., websites, reports) \\ 
\rowcolor{gray!30}
Written in English           & Non–peer-reviewed work  \\ 
Addresses LLMs in web accessibility   & LLM studies without accessibility focus  \\ 
\rowcolor{gray!30}
Focuses on web-based accessibility tasks   & Accessibility outside web context\\
Available in digital format  	& Full-text not available online\\

\hline 
\end{tabular}
}
\label{table:selection_criteria} 
\vspace{-0.2cm}
\end{table}

\textbf{Inclusion/Exclusion Criteria.}
Inclusion and exclusion criteria~\cite{patino2018inclusion} are essential for pruning the search space, reducing selection bias, and ensuring the retrieval of relevant, peer-reviewed scientific publications. Publications that meet these criteria serve as the starting point for manual filtering, enabling us to assess whether they align with the scope of this study, namely, the investigation of Large Language Models in web accessibility contexts. The resulting pool of publications also serves as the basis for backward snowballing~\cite{jalali2012systematic}, which involves examining the references cited by the selected studies, and forward snowballing~\cite{felizardo2016using}, which involves identifying studies that cite the selected publications. Table 2 summarizes the inclusion and exclusion criteria applied in this review.

Regarding the time range, we did not impose a starting date to capture early and foundational work in this emerging area. The end date for the search was set to November 30, 2025. Accordingly, the time range criterion allows the inclusion of any study published on or before this date, provided it satisfies the defined inclusion criteria.

\subsection{Execution}
The execution phase consists of conducting the literature search and applying a structured, multi-stage filtering process to identify the final set of primary studies. Figure \ref{fig:SearchProcess} provides an overview of this process and illustrates the number of publications retained at each stage. We executed the finalized search query across the six selected digital libraries, yielding an initial pool of 807 publications. To ensure rigor and consistency, all filtering stages were performed manually by multiple authors, with disagreements resolved through discussion until consensus was reached. 

In Stage 1, we removed duplicate records and retracted publications. This step reduced the initial pool from 807 to 752 candidate publications. In Stage 2, we applied the inclusion and exclusion criteria to the remaining publications' titles and abstracts. Studies that were clearly out of scope, for example, those that did not involve web accessibility or did not consider Large Language Models, were excluded at this stage. After title and abstract screening, 44 publications were retained for further analysis. In Stage 3, we conducted a full-text review of the remaining publications to assess their relevance in greater detail. During this stage, we excluded studies that did not substantively address LLMs in a web accessibility context or failed to meet the inclusion criteria upon closer inspection. This filtering step resulted in 31 publications. In Stage 4, we performed backward and forward snowball sampling on the selected studies by examining their reference lists and identifying papers that cited them. This process led to the identification of 2 additional relevant publications. After completing all filtering stages, we obtained a final set of 33 primary studies, which form the basis of the analysis presented in this review.

\subsection{Synthesis}
This phase synthesizes the extracted data to answer RQ$_1$–RQ$_5$. First, we classified the primary set of publications based on how Large Language Models are applied in web accessibility, distinguishing between studies that propose new LLM-based tools or systems and those that adopt or evaluate LLMs within existing accessibility workflows.

For RQ$_1$–RQ$_4$, we manually reviewed the full text of each publication and extracted concrete evidence regarding the accessibility tasks addressed, the use of accessibility standards and guidelines (e.g., WCAG and COGA), the types of accessibility issues targeted, and the technical design choices, including LLM models, prompting strategies, and system architectures. For RQ$_5$, we synthesized evaluation practices by examining the methods used and the involvement of participants, including users with disabilities.

All RQ-related data collected during the synthesis process were reviewed by two authors to mitigate bias, with disagreements resolved through discussion. The review was conducted by authors with prior research experience in web accessibility, supporting consistent interpretation during the selection and synthesis processes. A shared spreadsheet was used to document the extracted data and facilitate collaboration.

\section{Research Findings}
\label{sec:ResultsAndAnalysis}

This section presents the results of the review. The findings are organized by research question and summarize observed patterns in LLM applications, accessibility issues, and evaluation practices.





\begin{figure}[t]
\centering
\centering 
\begin{adjustbox}{width=0.5\textwidth,center}
\begin{tikzpicture}
\begin{axis}[
    xbar,
    width=\columnwidth,
    height=5.2cm,
    xmin=0,
    xmax=16,
    xtick distance=2,
    xlabel={Count},
    symbolic y coords={
        Automated Tool,
        Prompt Strategy,
        Browser Extension,
        Hybrid Approach,
        LLM Agent,
        Benchmark,
        Application
    },
    ytick=data,
    y dir=reverse,
    tick label style={font=\small},
    label style={font=\small},
    xmajorgrids=true,
    grid style={dotted},
    bar width=8pt,
    enlarge y limits=0.08,
    nodes near coords,
    nodes near coords align={horizontal},
    every node near coord/.append style={font=\tiny, xshift=2pt},
    axis x line*=bottom,
    axis y line*=left,
]
\addplot[
    fill=blue!60,
    draw=blue!90
] coordinates {
    (10,Automated Tool)
    (10,Prompt Strategy)
    (5,Browser Extension)
    (3,Hybrid Approach)
    (3,LLM Agent)
    (1,Benchmark)
    (1,Application)
};
\end{axis}
\end{tikzpicture}
\end{adjustbox}
\vspace{-0.6cm}
\caption{Distribution of solution types proposed in the reviewed studies.}
\label{fig:solution_types_clean}
\end{figure}
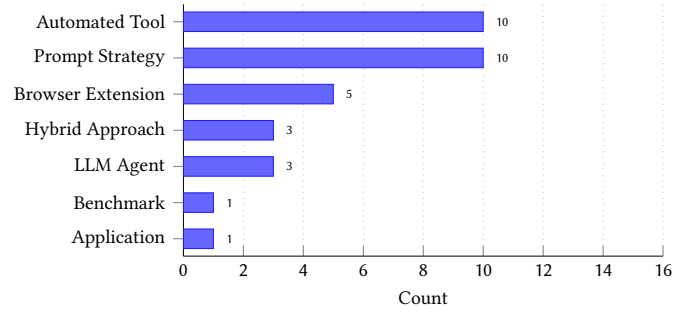

\vspace{-0.5cm}
\begin{box2}
\textbf{RQ$_1$: What are Large Language Models (LLMs) being applied to support web accessibility?}
\end{box2}
\vspace{-0.4cm}

Figure \ref{fig:solution_types_clean} presents the distribution of application types through which Large Language Models are used to support web accessibility in the reviewed studies. The results show that LLMs are most frequently applied as automated tools and through prompt-based strategies, each reported in 10 studies. These applications typically position LLMs as assistive components that support accessibility-related tasks, such as generating accessible content~\cite{panchanadikar2025can,li2025videoa11y,cheema2025describe,pedemonte2025improving,afsal2025websumm,moon2024sagol}, identifying accessibility issues~\cite{he2025enhancing,duarte2025expanding,paterno2025llm,andruccioli2025leveraging,lopez2025turning,abu2025can,acosta2024addressing}, or assisting with remediation workflows~\cite{aljedaani2025enhancing,fathallah2025accessguru,vera2025accessible,delnevo2024interaction,othman2023fostering}. Browser extensions represent the next most common application type, appearing in 5 studies, reflecting efforts to integrate LLM-based accessibility support directly into development or browsing environments~\cite{yu2025cluttered,pedemonte2025improving,afsal2025websumm,silva2024page,huynh2024smartcaption}.

Less frequently, studies apply LLMs within hybrid approaches and LLM agent–based systems, each reported in 3 studies, indicating limited adoption of more complex or autonomous architectures~\cite{fathallah2025accessguru,anderer2025making,lopez2025turning}. Only one study frames LLMs as part of an explicit benchmark~\cite{andruccioli2025tabular}, and one study presents a standalone application~\cite{moon2024sagol}, suggesting that relatively few works focus on systematic evaluation infrastructures or end-to-end accessibility systems. An examination of the study contexts in Table \ref{tab:study_summary} shows that most applications are evaluated in controlled or task-specific settings, such as standardized prompting, limited datasets, or individual web pages, rather than in long-term or real-world deployment scenarios. This indicates that current applications primarily emphasize feasibility and task performance over sustained integration into production-level accessibility workflows.



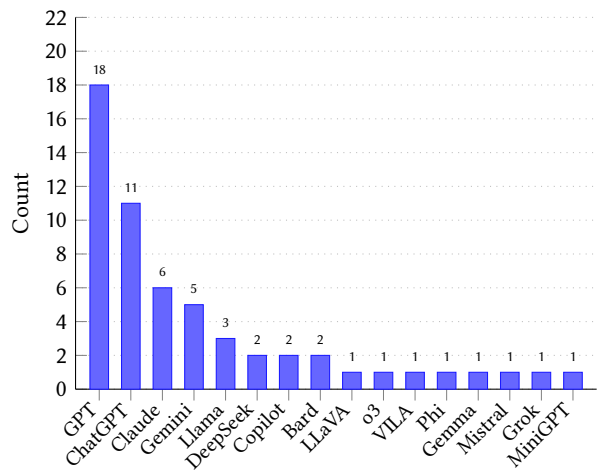
\begin{figure}[t]
\centering
\begin{tikzpicture}
\begin{axis}[
    ybar,
    width=\linewidth,
    height=6.5cm,
    ymin=0,
    ymax=22,
    ytick distance=2,
    ylabel={Count},
    symbolic x coords={
        GPT,ChatGPT,Claude,Gemini,Llama, DeepSeek,
        Copilot,Bard, LLaVA, o3,VILA,Phi,Gemma,Mistral,Grok,MiniGPT
    },
    xtick=data,
    xticklabel style={rotate=45, anchor=east, font=\small},
    ymajorgrids=true,
    grid style={dotted},
    bar width=7pt,
    enlarge x limits=0.05,
    nodes near coords,
    nodes near coords align={vertical},
    every node near coord/.append style={font=\tiny, yshift=1pt},
    axis x line*=bottom,
    axis y line*=left,
]

\addplot[
    fill=blue!60,
    draw=blue!90
] coordinates {
    (GPT,18)
    (ChatGPT,11)
    (Claude,6)
    (Gemini,5)
    (Llama,3)
    (DeepSeek,2)
    (Copilot,2)
    (Bard,2)
    (o3,1)
    (LLaVA,1)
    (VILA,1)
    (Phi,1)
    (Gemma,1)
    (Mistral,1)
    (Grok,1)
    (MiniGPT,1)
};

\end{axis}
\end{tikzpicture}
\vspace{-0.4cm}
\caption{Distribution of large language models (LLMs) reported in the reviewed studies.}
\label{fig:llm_distribution_barchart}
\end{figure}

\begin{figure}[t]
\centering
\begin{tikzpicture}
\begin{axis}[
    ybar,
    width=\columnwidth,
    height=6.2cm,
    ymin=0,
    ymax=14,
    ytick distance=2,
    ylabel={\scriptsize Count},
    symbolic x coords={
        Not reported,
        Instruction,
        Role,
        Zero-shot,
        Structured,
        Verbosity,
        Iterative,
        Self-Reflection,
        Agent/Tool-Use,
        Sequential,
        Few-shot,
        One-shot,
        Custom/Specific
    },
    xtick=data,
    xticklabel style={
        rotate=45,
        anchor=east,
        font=\scriptsize
    },
    tick label style={font=\scriptsize},
    ymajorgrids=true,
    grid style={dotted},
    bar width=6pt,
    enlarge x limits=0.10,
    nodes near coords,
    nodes near coords align={vertical},
    every node near coord/.append style={font=\tiny, yshift=1pt},
    axis x line*=bottom,
    axis y line*=left,
]

\addplot[
    fill=blue!60,
    draw=blue!90
] coordinates {
    (Not reported,13)
    (Instruction,10)
    (Role,9)
    (Zero-shot,6)
    (Structured,6)
    (Verbosity,6)
    (Iterative,4)
    (Self-Reflection,2)
    (Agent/Tool-Use,2)
    (Sequential,1)
    (Few-shot,1)
    (One-shot,1)
    (Custom/Specific,1)
};

\end{axis}
\end{tikzpicture}
\caption{Counts of prompting techniques across the reviewed papers.}
\label{fig:prompting_techniques_bar}
\end{figure}
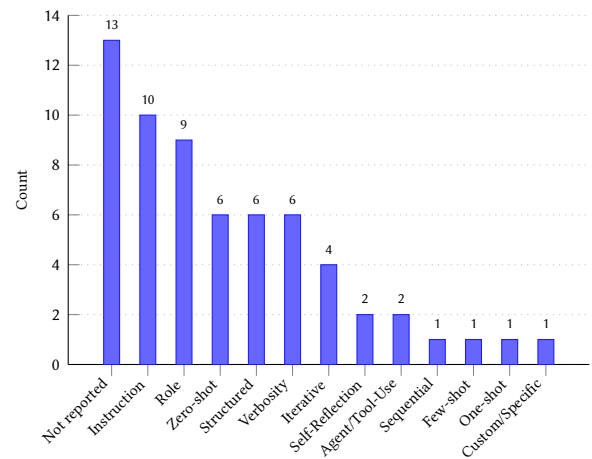


\vspace{-0.5cm}
\begin{box2}
\textbf{RQ$_3$: What LLM models, prompting strategies, and system architectures are used in web accessibility research?}
\end{box2}
\vspace{-0.4cm}


Figure \ref{fig:llm_distribution_barchart} shows the distribution of Large Language Models used across the reviewed studies. The results reveal a strong concentration around commercial, general-purpose LLMs, particularly GPT-based models. GPT-4 is the most frequently used model, appearing in 18 studies, followed by ChatGPT in 11 studies and Claude in 6 studies. Other models, including Gemini, LLaMA, DeepSeek, and Copilot, are used less frequently, while vision–language and multimodal models such as LLaVA, VILA, Phi, MiniGPT appear only sporadically. This distribution indicates that most web accessibility research prioritizes readily available, high-capacity LLMs rather than models specifically trained or fine-tuned for accessibility-related tasks.

A closer examination of Table \ref{tab:study_summary} shows that these models are primarily employed in comparative or head-to-head evaluation settings, where multiple LLMs are tested on the same accessibility task or dataset. In several studies, model performance is assessed relative to baseline tools, human experts, or alternative LLMs, suggesting that model choice is often treated as a key experimental variable. However, only a limited number of studies report systematic exploration of model configuration, fine-tuning, or training data characteristics, indicating that accessibility outcomes are largely evaluated at the level of out-of-the-box model behavior.

Figure \ref{fig:prompting_techniques_bar} summarizes the prompting strategies adopted across the reviewed studies. The most common approaches are instruction-based prompting (10 studies) and role-based prompting (9 studies), followed by zero-shot, structured, and verbosity-controlled prompting (each reported in 6 studies). These strategies are typically used to guide LLMs toward producing accessibility-relevant outputs, such as emphasizing compliance with WCAG success criteria or generating detailed descriptions. More advanced prompting techniques—such as self-reflection, agent-based prompting, ensemble prompting, and chain-of-thought-like strategies—are comparatively rare, appearing in only one or two studies each.

When considered alongside the system descriptions in Table \ref{tab:study_summary}, these findings indicate that most studies adopt relatively simple system architectures, in which a single LLM is invoked via carefully crafted prompts to perform specific accessibility tasks. Only a small subset of studies employ multi-agent systems~\cite{fathallah2025accessguru}, iterative feedback loops~\cite{gurițua2025exploring}, or hybrid pipelines~\cite{paterno2025llm} that combine LLMs with rule-based tools or accessibility checkers. Overall, the results indicate that current research places greater emphasis on model selection and prompt design than on architectural innovation, with limited exploration of persistent, adaptive, or large-scale accessibility systems built around LLMs

\begin{table*}[t]
\centering
\caption{Mapping of primary studies (PS) to accessibility guidelines. A checkmark indicates that the guideline is referenced in the study.}
\vspace{-0.4cm}
\label{tab:ps_guidelines_matrix}
\begin{adjustbox}{width=0.85\textwidth,center}

\begin{tabular}{|c|c|c|c|c|c|c|c|c|c|c|c|c|c|c|}
\hline
\rowcolor{gray!60}
\textbf{PS} &
\textbf{WCAG 2.0} &
\textbf{WCAG 2.1} &
\textbf{WCAG 2.2} &
\textbf{WAI-ARIA} &
\textbf{Ofcom} &
\textbf{EN 301 549} &
\textbf{COGA} &
\textbf{HTML API} &
\textbf{Netflix} &
\textbf{MAC} &
\textbf{DCMP} &
\textbf{Section 508} &
\textbf{NCAM} &
\textbf{Not Reported} \\
\hline

PS\#1 \cite{panchanadikar2025can}  &  &  & \cmark &  &  &  & \cmark &  &  &  &  &  &  &  \\
\rowcolor{gray!30}
PS\#2 \cite{he2025enhancing}  &  & \cmark & \cmark & \cmark &  &  &  & \cmark &  &  &  &  &  &  \\
PS\#3 \cite{aljedaani2025enhancing}  & \cmark &  &  &  &  &  &  &  &  &  &  &  &  &  \\
\rowcolor{gray!30}
PS\#4 \cite{duarte2025expanding}  &  &  & \cmark &  &  &  &  &  &  &  &  &  &  &  \\
PS\#5 \cite{gurițua2025exploring}  &  & \cmark &  &  &  &  &  &  &  &  &  &  &  &  \\
\rowcolor{gray!30}
PS\#6 \cite{leedy2025accessibility}  &  & \cmark &  &  &  &  &  &  &  &  &  &  &  &  \\
PS\#7 \cite{yu2025cluttered}  &  & \cmark &  &  &  &  &  &  &  &  &  &  &  &  \\
\rowcolor{gray!30}
PS\#8 \cite{gurita2025llm}  &  & \cmark &  &  &  &  &  &  &  &  &  &  &  &  \\
PS\#9 \cite{paterno2025llm}  & \cmark &  &  &  &  &  &  &  &  &  &  &  &  &  \\
\rowcolor{gray!30}
PS\#10 \cite{fathallah2025accessguru} &  & \cmark &  &  &  &  &  &  &  &  &  &  &  &  \\
PS\#11 \cite{doush2024evaluating} &  & \cmark &  &  &  &  &  &  &  &  &  &  &  &  \\
\rowcolor{gray!30}
PS\#12 \cite{cao2025scenegena11y} & \cmark &  &  &  &  &  &  &  &  &  &  &  &  &  \\

PS\#13 \cite{anderer2025making} &  &  &  &  &  &  &  &  &  &  &  &  &  & \cmark \\
\rowcolor{gray!30}
PS\#14 \cite{li2025videoa11y} &  &  &  &  & \cmark &  &  &  & \cmark & \cmark & \cmark &  &  &  \\

PS\#15 \cite{mo2025tablenarrator} & \cmark &  &  &  &  &  &  &  &  &  &  &  &  &  \\
\rowcolor{gray!30}
PS\#16 \cite{cheema2025describe} & \cmark &  &  &  & \cmark &  &  &  &  &  &  &  &  &  \\

PS\#17 \cite{vera2025accessible} &  &  & \cmark &  &  &  &  &  &  &  &  &  &  &  \\
\rowcolor{gray!30}
PS\#18 \cite{pedemonte2025improving} & \cmark &  &  &  &  &  &  &  &  &  &  &  & \cmark &  \\
PS\#19 \cite{andruccioli2025leveraging} &  & \cmark &  &  &  &  &  &  &  &  &  &  &  &  \\
\rowcolor{gray!30}
PS\#20 \cite{andruccioli2025tabular} &  & \cmark &  &  &  &  &  &  &  &  &  &  &  &  \\
PS\#21 \cite{lopez2025turning} & \cmark &  &  &  &  &  &  &  &  &  &  &  &  &  \\
\rowcolor{gray!30}
PS\#22 \cite{afsal2025websumm} & \cmark &  &  &  &  &  &  &  &  &  &  &  &  &  \\

PS\#23 \cite{michalichem2025gamified} & \cmark &  &  &  &  &  &  &  &  &  &  &  &  &  \\
\rowcolor{gray!30}
PS\#24 \cite{abu2025can} &  & \cmark &  &  &  &  &  &  &  &  &  &  &  &  \\

PS\#25 \cite{afsal2024comparative} & \cmark &  &  &  &  &  &  &  &  &  &  &  &  &  \\
\rowcolor{gray!30}
PS\#26 \cite{singh2024accessibility} & \cmark &  &  &  &  &  &  &  &  &  &  &  &  &  \\
PS\#27 \cite{silva2024page} & \cmark &  &  & \cmark &  &  &  &  &  &  &  &  &  &  \\
\rowcolor{gray!30}
PS\#28 \cite{delnevo2024interaction} & \cmark &  &  &  &  &  &  &  &  &  &  &  &  &  \\
PS\#29 \cite{huynh2024smartcaption} & \cmark &  &  &  &  &  &  &  &  &  &  &  &  &  \\
\rowcolor{gray!30}
PS\#30 \cite{oswal2024examining} &  & \cmark &  &  &  & \cmark &  &  &  &  &  & \cmark &  &  \\
PS\#31 \cite{acosta2024addressing} &  &  & \cmark &  &  &  &  &  &  &  &  &  &  &  \\
\rowcolor{gray!30}
PS\#32 \cite{moon2024sagol} &  & \cmark &  &  &  &  &  &  &  &  &  &  &  &  \\
PS\#33 \cite{othman2023fostering} &  & \cmark &  &  &  &  &  &  &  &  &  &  &  &  \\
\hline
\end{tabular}
\end{adjustbox}

\end{table*}

\vspace{-0.5cm}
\begin{box2}
\textbf{RQ$_4$: What web accessibility issues and WCAG success-criterion violations are most frequently addressed?}
\end{box2}
\vspace{-0.4cm}


\begin{figure}[t]
\centering
\begin{tikzpicture}
\begin{axis}[
    ybar,
    width=\columnwidth,
    height=6.2cm,
    ymin=0,
    ymax=14,
    ytick distance=2,
    ylabel={Count},
    symbolic x coords={
        Empirical,
        Qualitative,
        {Case Study},
        Experimental,
        {Mixed-Methods},
        {User-Centered Study},
        Exploratory,
        Comparative,
        {Empirical \& Case Study},
        {Preliminary Eval.}
    },
    xtick=data,
    xticklabel style={rotate=45, anchor=east, font=\small},
    tick label style={font=\small},
    ymajorgrids=true,
    grid style={dotted},
    bar width=6pt,
    enlarge x limits=0.12,
    nodes near coords,
    nodes near coords align={vertical},
    every node near coord/.append style={font=\tiny, yshift=1pt},
    axis x line*=bottom,
    axis y line*=left,
]

\addplot[
    fill=blue!60,
    draw=blue!90
] coordinates {
    (Empirical,12)
    (Qualitative,4)
    ({Case Study},3)
    (Experimental,4)
    ({Mixed-Methods},3)
    ({User-Centered Study},2)
    (Exploratory,2)
    (Comparative,1)
    ({Empirical \& Case Study},1)
    ({Preliminary Eval.},1)
};

\end{axis}
\end{tikzpicture}

\vspace{-0.4cm}
\caption{Distribution of study types across the reviewed papers (N=36).}
\label{fig:study_types_vertical}
\end{figure}
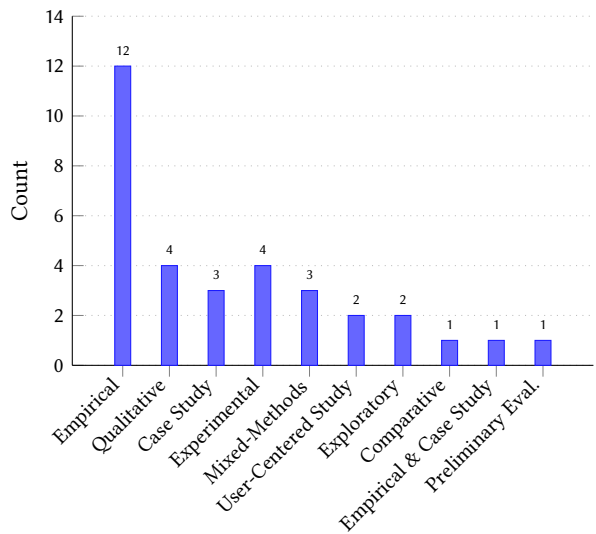

\begin{figure}[t]
\centering
\begin{adjustbox}{width=0.35\textwidth,center}
\begin{tikzpicture}

\pie[
    radius=2.1,
    text=legend,
    sum=auto,
    after number={\%}
]{
    46.51/Visual Impairment,
    20.93/{General/Unspecified},
    13.95/{Cognitive/Intellectual},
    9.30/{Deaf/HoH},
    6.98/{Motor/Physical},
    2.33/Not Reported
}

\fill[white] (0,0) circle (1.25);

\node[font=\small] at (0,0) {N=43};

\end{tikzpicture}
\end{adjustbox}

\vspace{-0.4cm}
\caption{Distribution of disability types reported in the reviewed studies (N=33).}
\label{fig:disability_donut}
\end{figure}
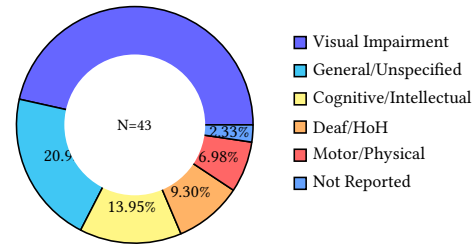
\begin{table}[t]
\centering
\caption{Participant types and total number of participants reported in the primary studies (PS).}
\vspace{-0.4cm}
\label{tab:participants_verbatim}
\resizebox{\columnwidth}{!}{%
\begin{tabular}{|c|p{9.5cm}|c|}
\hline
\rowcolor{gray!60}
\textbf{Study} & \textbf{Participant Types} & \textbf{\# Total} \\
\hline

PS\#2 \cite{he2025enhancing}  & Authors & 2 \\ \hline

PS\#3 \cite{aljedaani2025enhancing}  & Students & 215 \\ \hline
\rowcolor{gray!15}
PS\#5 \cite{gurițua2025exploring}  & Designers/UX Professionals & 6 \\ \hline

PS\#6 \cite{leedy2025accessibility}  & Reviewers/Evaluators & 2 \\ \hline

\rowcolor{gray!15}
PS\#7 \cite{yu2025cluttered}  & Blind (17), Low Vision Users (4) & 21 \\ \hline

PS\#8 \cite{gurita2025llm}  & Accessibility Expert & 2 \\ \hline

\rowcolor{gray!15}
PS\#9 \cite{paterno2025llm}  & Accessibility Expert & 1 \\ \hline


PS\#12 \cite{cao2025scenegena11y} & Blind/Low Vision Users (3), Deaf/Hard of Hearing Users (3) & 6 \\ \hline

\rowcolor{gray!15}
PS\#13 \cite{anderer2025making} & Students & 12 \\ \hline

PS\#14 \cite{li2025videoa11y}  & Blind/Low Vision Users(40), Professional Describers (7), Sighted Users(347) & 394 \\ \hline

\rowcolor{gray!15}
PS\#15 \cite{mo2025tablenarrator} & Blind/Low Vision Users(8), Reviewers/Evaluators (5), Sighted Users(11) & 24 \\ \hline

PS\#16 \cite{cheema2025describe} & Blind (15) Low Vision Users (5) & 20 \\ \hline


\rowcolor{gray!15}
PS\#17 \cite{vera2025accessible} & Low Vision Users (2), Sighted Users(2) & 4 \\ \hline

PS\#18 \cite{pedemonte2025improving} & Low Vision Users (1), Students (34) & 35 \\ \hline

\rowcolor{gray!15}
PS\#19 \cite{andruccioli2025leveraging} & Developers/Engineers (88), Students(127) & 215 \\ \hline


PS\#22 \cite{afsal2025websumm} & Low Vision Users & 22 \\ \hline

\rowcolor{gray!15}
PS\#23 \cite{michalichem2025gamified} & Reviewers/Evaluators(3), Students(10) & 13 \\ \hline

PS\#27 \cite{silva2024page} & Blind (12), Low Vision Users (3) & 15 \\ \hline

\rowcolor{gray!15}
PS\#29 \cite{huynh2024smartcaption} & General Participants & 3 \\ \hline

PS\#30 \cite{oswal2024examining} & Authors & 2 \\ \hline

\rowcolor{gray!15}
PS\#31 \cite{acosta2024addressing} & Accessibility Experts & 5 \\ \hline














\rowcolor{gray!15}
\multicolumn{2}{|l|}{\textit{\textbf{No participant information reported in} PS\#1, PS\#10, PS\#11, PS\#4, PS\#32, PS\#33,
}} & 0 \\
\rowcolor{gray!15}
\multicolumn{2}{|l|}{\textit{ PS\#24, PS\#25, PS\#26, PS\#28, PS\#20, PS\#21}} &  \\ \hline

\end{tabular}
}
\end{table}

Overall, the reviewed literature demonstrates a strong concentration on a limited set of accessibility issues, primarily those related to content perceivability and textual representation, with most studies explicitly grounding their analyses in established accessibility frameworks such as the Web Content Accessibility Guidelines (WCAG) and, to a lesser extent, the Cognitive Accessibility Guidelines (COGA). As summarized in Table \ref{tab:ps_guidelines_matrix}, WCAG serves as the dominant reference framework for defining, detecting, and evaluating accessibility issues across the reviewed studies, while COGA guidelines appear in a smaller subset, reflecting a more limited but emerging engagement with cognitive accessibility concerns.

Within this framing, the most frequently addressed issue is missing or vague alternative text for non-text content, which appears in 21 studies (Table \ref{tab:accessibility_issues_complete}), making it the most dominant accessibility problem targeted by LLM-based approaches. This emphasis aligns closely with LLMs' strengths in natural language generation and refinement. A second prominent cluster of issues related to visual presentation and perceivability, particularly insufficient color contrast, which is reported in 16 studies and is often treated as a detectable, correctable violation. Together, these findings indicate that current research prioritizes accessibility issues that can be framed as missing textual equivalents or visually measurable deficiencies.

Beyond perceivability, a substantial portion of the literature focuses on structural, semantic, and programmatic accessibility issues that affect how web content is interpreted by assistive technologies. Commonly addressed problems include improper heading structure, missing or insufficient labels and instructions, incorrect form labeling or association, and inaccessible name, role, or value information, each appearing across multiple studies. These issues highlight recurring deficiencies in how semantic relationships and interaction cues are encoded in web interfaces, particularly for screen reader users. In addition, keyboard navigation and focus management issues are addressed in 11 studies (Table \ref{tab:accessibility_issues_complete}), indicating moderate attention to operability barriers faced by users who rely on non-pointer input modalities. However, other navigation-related concerns—such as missing bypass mechanisms~\cite{yu2025cluttered,vera2025accessible,silva2024page}, ambiguous link purposes~\cite{leedy2025accessibility,lopez2025turning,singh2024accessibility}, or inconsistent focus order—are addressed less frequently~\cite{cao2025scenegena11y}, suggesting uneven coverage even within interaction-focused accessibility research.

In contrast, cognitive accessibility and language-related issues, as captured by COGA guidelines, receive comparatively limited attention in the reviewed literature. Only a small number of studies~\cite{panchanadikar2025can,aljedaani2025enhancing,michalichem2025gamified,abu2025can,afsal2024comparative,singh2024accessibility} address issues such as content overload, long uninterrupted text, lack of clear structure, or support for undoing actions, despite their known impact on users with cognitive and learning disabilities. Similarly, issues related to page language identification, reading level, and time-based media accessibility (e.g., captions and audio descriptions) appear sporadically. The table also reveals a set of LLM-specific and emergent accessibility issues that do not map cleanly to individual WCAG success criteria, including hallucinated image descriptions~\cite{li2025videoa11y,huynh2024smartcaption,moon2024sagol}, dataset or annotation quality problems~\cite{gurițua2025exploring,li2025videoa11y,andruccioli2025tabular,oswal2024examining}, broken or incorrect ARIA usage~\cite{leedy2025accessibility,andruccioli2025leveraging,oswal2024examining,acosta2024addressing}, and design flaws introduced during automated generation~\cite{panchanadikar2025can,leedy2025accessibility,gurita2025llm}. Although less frequent, these issues are particularly significant in the context of LLM-based systems, as they represent failure modes introduced or amplified by generative technologies, underscoring that accessibility research involving LLMs must address both long-standing web accessibility barriers and new risks associated with automated content generation.


\begin{table*}[t]
\centering
\caption{Accessibility issues and violations identified across the reviewed primary studies.}
\vspace{-0.4cm}
\label{tab:accessibility_issues_complete}
\resizebox{\textwidth}{!}{%
\begin{tabular}{|l|p{5.3cm}|p{5.3cm}|c||l|p{5.3cm}|p{5.3cm}|c|}
\hline
\rowcolor{gray!60}
\textbf{Code} & \textbf{Issue / Violation} & \textbf{Primary Studies (PS)} & \textbf{Cnt} &
\textbf{Code} & \textbf{Issue / Violation} & \textbf{Primary Studies (PS)} & \textbf{Cnt} \\
\hline
WCAG SC 1.1.1 &
Non-text content missing or vague alternative text &
PS\#1, PS\#2, PS\#3, PS\#6, PS\#7, PS\#8, PS\#9, PS\#10, PS\#13, PS\#15, PS\#17, PS\#18, PS\#20, PS\#21, PS\#26, PS\#28, PS\#29, PS\#30, PS\#31, PS\#32, PS\#33 &
21 &
WCAG SC 1.2.2 &
Missing or insufficient captions (prerecorded) &
PS\#10 &
1 \\

\rowcolor{gray!30}
WCAG SC 1.3.1 &
Improper heading structure &
PS\#2, PS\#4, PS\#6, PS\#7, PS\#9, PS\#20, PS\#24, PS\#26, PS\#27, PS\#30, PS\#33 &
11 &
WCAG SC 1.3.1 &
Incorrect form labeling or association &
PS\#1, PS\#3, PS\#8, PS\#11 &
4 \\

WCAG SC 1.3.1 &
Semantic structure issues &
PS\#17, PS\#20, PS\#31  &
3 &
WCAG SC 1.4.1 &
Color used as the sole means of conveying information &
PS\#8 &
1 \\

\rowcolor{gray!30}
WCAG SC 1.4.3 &
Insufficient color contrast (minimum) &
PS\#1, PS\#2, PS\#3, PS\#5, PS\#6, PS\#8, PS\#10, PS\#12, PS\#13, PS\#17, PS\#19, PS\#21, PS\#26, PS\#30, PS\#31, PS\#33 &
16 &
WCAG SC 1.4.4 &
Resize text not supported &
PS\#3, PS\#8, PS\#13  &
3 \\

WCAG SC 1.4.6 &
Contrast (enhanced) not met &
PS\#3 &
1 &
WCAG SC 1.4.8 &
Visual presentation issues (brightness/clutter) &
PS\#12 &
1 \\

\rowcolor{gray!30}
WCAG SC 1.4.12 &
Text spacing issues &
PS\#10 &
1 &
WCAG SC 2.1.1 &
Keyboard navigation / tabindex issues &
PS\#6, PS\#8, PS\#10, PS\#11, PS\#13, PS\#17, PS\#20, PS\#24, PS\#30, PS\#31, PS\#33 &
11 \\

WCAG SC 2.4.1 &
Missing navigation aids (bypass blocks) &
PS\#7, PS\#17, PS\#27 &
3 &
WCAG SC 2.4.2 &
Missing or incorrect page titles &
PS\#3 &
1 \\

\rowcolor{gray!30}
WCAG SC 2.4.3 &
Incorrect focus order &
PS\#12 &
1 &
WCAG SC 2.4.4 &
Non-descriptive link text &
PS\#2, PS\#3, PS\#10, PS\#33 &
4 \\

WCAG SC 2.4.4 &
Ambiguous link purpose &
PS\#6, PS\#21, PS\#26 &
3 &
WCAG SC 2.4.6 &
Unclear or insufficient headings and labels &
PS\#1, PS\#3, PS\#8, PS\#12  &
4 \\

\rowcolor{gray!30}
WCAG SC 2.4.7 &
Focus not visible &
PS\#8, PS\#11 &
2 &
WCAG SC 2.5.3 &
Label not included in accessible name &
PS\#8 &
1 \\

WCAG SC 2.5.5 &
Insufficient target size &
PS\#5, PS\#8, PS\#10 &
3 &
WCAG SC 3.1.1 &
Missing or incorrect page language &
PS\#3, PS\#8, PS\#21 &
3 \\

\rowcolor{gray!30}
WCAG SC 3.1.2 &
Language of parts not identified &
PS\#10 &
1 &
WCAG SC 3.1.5 &
Reading level issues (AAA) &
PS\#25 &
1 \\

WCAG SC 3.2.4 &
Inconsistent identification &
PS\#18 &
1 &
WCAG SC 3.3.1 &
Error identification missing &
PS\#23 &
1 \\

\rowcolor{gray!30}
WCAG SC 3.3.2 &
Missing or insufficient labels or instructions &
PS\#1, PS\#2, PS\#3, PS\#9, PS\#10, PS\#17, PS\#20, PS\#26, PS\#28, PS\#30, PS\#33 &
11 &
WCAG SC 4.1.1 &
Parsing errors &
PS\#3, PS\#8  &
2 \\

WCAG SC 4.1.2 &
Inaccessible name, role, or value &
PS\#2, PS\#3, PS\#7, PS\#8, PS\#10, PS\#12, PS\#15, PS\#17, PS\#19, PS\#28, PS\#30, PS\#32 &
12 &
WCAG SC 4.1.3 &
Missing or improper status messages &
PS\#8, PS\#13, PS\#19 &
3 \\

\rowcolor{gray!30}
WCAG 2.2 &
Adjacent links pointing to same destination &
PS\#1 &
1 &
WCAG SC 1.2.5 &
Missing audio description (prerecorded) &
PS\#12, PS\#16, PS\#29 &
3 \\

COGA &
Clearly identify controls and their use &
PS\#1 &
1 &
COGA &
Lack of white space / long uninterrupted text &
PS\#1 &
1 \\

\rowcolor{gray!30}
COGA &
Let users go back or undo actions &
PS\#1 &
1 &
COGA &
Make each step clear &
PS\#1 &
1 \\

COGA &
Missing information about data or fees &
PS\#1 &
1 &
COGA &
No icons accompanying headings &
PS\#1 &
1 \\

\rowcolor{gray!30}
COGA &
Static or non-functional buttons and unclear layouts &
PS\#1 &
1 &
COGA &
Clear and understandable page structure missing &
PS\#1 &
1 \\

Other &
Hallucinated image descriptions or objects &
PS\#13, PS\#15, PS\#18, PS\#29 &
4 &
Other &
Broken ARIA references &
PS\#33 &
1 \\

\rowcolor{gray!30}
Other &
Content overload / linear navigation burden &
PS\#22, PS\#27 &
2 &
Other &
Dataset or annotation quality issues &
PS\#14, PS\#16, PS\#18, PS\#29 &
4 \\

Other &
Design flaws or broken layout &
PS\#24 &
1 &
Other &
HTML / CSS validation errors &
PS\#23, PS\#26 &
2 \\

\hline
\end{tabular}
}
\end{table*}

\vspace{-0.6cm}
\begin{box2}
\textbf{RQ$_5$: How are LLM-based web accessibility approaches evaluated?}
\end{box2}
\vspace{-0.5cm}


Figure \ref{fig:disability_donut} summarizes the evaluation methods employed across the reviewed studies and shows that evaluation practices are dominated by empirical and inspection-based approaches. Empirical evaluations are the most common, appearing in 12 studies, followed by qualitative evaluations (4 studies), case studies (3 studies), and experimental designs (4 studies). A comparable number of studies employ mixed-methods or user-centered evaluations (3 studies each), while fewer works rely on exploratory, comparative, or preliminary evaluations. This distribution indicates that most studies seek to demonstrate feasibility or performance through controlled experiments, structured inspections, or proof-of-concept deployments, rather than through long-term field studies or large-scale longitudinal evaluations.

Table \ref{tab:participants_verbatim} further details the extent and nature of participant involvement in these evaluations. While several studies include human participants, participant-based evaluations vary widely in scale and composition. Many studies rely on authors, students, or developers as participants, often as proxies for end users. A smaller subset explicitly involves accessibility experts, reviewers, or users with disabilities, including blind, low-vision, deaf, or hard-of-hearing participants. Even when users with disabilities are included, sample sizes are typically modest and heterogeneous, with studies often combining multiple participant groups within a single evaluation. Notably, a non-trivial number of studies report no participant information at all, relying instead on automated audits, benchmark datasets, or expert inspection without direct user involvement.

Taken together, these results show that evaluation of LLM-based web accessibility approaches is methodologically diverse but uneven. While many studies employ empirical or experimental designs, direct validation with people with disabilities remains limited and inconsistent. The evidence base is therefore shaped largely by tool-centric and developer-centric evaluations, with comparatively fewer studies providing user-centered or participatory validation. This pattern suggests that current evaluation practices emphasize technical feasibility and comparative performance, while real-world effectiveness and user experience are less systematically assessed.

\begin{table*}
\centering
\caption{Summary of study contexts, experimental settings, models, and reported results across primary studies.}
\vspace{-0.4cm}
\label{tab:study_summary}
\scriptsize
\resizebox{\textwidth}{!}{%

\renewcommand{\arraystretch}{1.15}
\begin{tabular}{|p{0.4cm}|p{2.5cm}|p{2.5cm}|p{1.5cm}|p{1.5cm}|p{1.0cm}|p{1.2cm}|p{7.0cm}|}
\hline
\rowcolor{gray!60}
\textbf{Study} &
\textbf{Objective} &
\textbf{Experiment / Setting} &
\textbf{Website Type} &
\textbf{Models} &
\textbf{Best Model} &
\textbf{Comparison Type} &
\textbf{Result by Model} \\
\hline

PS\#1 \cite{panchanadikar2025can} &
Evaluate WCAG 2.2 compliance of AI-generated websites \& compare ChatGPT vs. DeepSeek accessibility errors &
Standardized prompting of LLMs to generate HTML/CSS websites, then evaluated for WCAG and cognitive &
E-commerce (online clothing store) &
DeepSeek, ChatGPT &
DeepSeek &
head-to-head &
ChatGPT: Generated 54.05\% of errors (140 errors); fewer errors in Buttons (38.24\% vs 61.76\%) and Misc Content (44.62\% vs 55.38\%). DeepSeek: Generated 45.95\% of errors (119 errors); fewer errors in Headings (29.63\% vs 70.37\%) and Forms (34.21\% vs 65.79\%). \\
\hline

\rowcolor{gray!15}
PS\#2 \cite{he2025enhancing} &
Introduce GENA11Y, an automated accessibility checker to detect WCAG violations &
GenA11y tool vs existing tools and ablation study on datasets and real websites &
Real websites (Dataset Dr) and datasets (Da, Dm) &
GPT-4o &
GPT-4o &
GenA11y vs existing tools &
Experiment: GENA11Y (GPT-4o): 87.61\% recall (516 issues); Next best tool (IBM): 38.20\% recall. 
Abalation: Base Model (LLM with raw HTML): Detected 57 issues; GENA11Y: Detected 127 issues. \\
\hline

PS\#3 \cite{aljedaani2025enhancing} &
Integrate LLM into software eng. courses to improve web accessibility skills.  &
Participants used ChatGPT to fix/repair identified issues &
Participant websites &
ChatGP 3.5 &
ChatGP 3.5 &
ChatGP 3.5 vs WAVE &
ChatGP 3.5: Fixed 47\% (AChecker assessment) and 46\% (WAVE assessment) of violations. \\
\hline

\rowcolor{gray!15}
PS\#4 \cite{duarte2025expanding} &
Detecting heading-related accessibility barriers  &
Models answered 6 A11y questions for 10 webpages  &
W3C webpage &
Llama 3.1, GPT-4o, GPT-4o mini &
GPT-4o mini &
head-to-head &
GPT-4o mini: Accuracy 73.33\%, F-measure 82.61\%. Llama 3.1: Accuracy 70.00\%, F-measure 79.54\%. GPT-4o: Accuracy 70.00\%, F-measure 79.07\%. \\
\hline

PS\#5 \cite{gurițua2025exploring} &
Framework for Human-AI collaboration in UI accessibility via compliance \& prompt eng. &
Evaluated 50 UIs from 5 AI design tools; tested 4 tools on bad design tasks; and 200 UIs &
Not reported &
Not explicitly named &
Not explicitly named &
N/A &
AI design tools (aggregate): "modest violations (M=0.47 on an 0-4 scale)". AI design tools (bad design task): Showed "significant resistance". Iterative prompting: Improved through dialogue in 42\% of cases. \\
\hline

\rowcolor{gray!15}
PS\#6 \cite{leedy2025accessibility} &
Assess WCAG compliance of Wix and Framer AI website builders &
Comparison of AI website builders (Wix vs Framer) using standardized prompts &
E-commerce, Healthcare, School for Blind &
Not reported &
Wix &
tool vs tool &
Wix: Average errors M = 33.6 (SD = 10.5); Provided alt text in 100\% of cases. Framer: Average errors M = 63.8 (SD = 12.2); Failed to provide alternative text in 100\% of cases. \\
\hline

PS\#7 \cite{yu2025cluttered} &
To restructure HTML for accessible websites, validated with screen reader users &
User study and automated auditing of extension variants vs original websites &
E-commerce (Mercari, Amazon, Nordstrom) &
GPT-4o &
GPT-4o &
baseline vs LLM &
Option 1 regenerates HTML to enforce logical order, while Option 2 reorganizes existing tags. Both reduced task time (p < 0.001), with Option 1 performing better (p < 0.001). \\
\hline

\rowcolor{gray!15}
PS\#8 \cite{gurita2025llm} &
Compare ChatGPT and Claude on WCAG 2.1 compliance using accessibility-oriented prompts &
Comparison of two LLMs using accessibility-agnostic and accessibility-oriented prompts, plus self-correction &
Banking app homepages &
Claude 3.5 Haiku, GPT-4-turbo &
Tie &
head-to-head | prompt variant &
Expert Evaluation: Severity of issues (0-4 scale). Claude 3.5 Haiku consistently produced UIs with lower severity issues than ChatGPT across both prompt types (0.6 vs 1.5 agnostic; 0.2 vs 0.4 oriented). Both models successfully reduced the severity of accessibility issues to 0.0 after the self-reflection step. \\
\hline

PS\#9 \cite{paterno2025llm} &
Investigate GPT-4 to automate validation of manual WCAG techniques &
Comparison of Manual Testing vs Hybrid Testing (Manual + GenAI) &
Travel website (skyscanner.net clone) &
GPT-4o &
GPT-4o &
N/A &
GPT-4o produced 251 responses, of which 232 were correct (92.4\%): 185 successes, 54 failures, and 12 warnings. Across 45 images, TeVal/GPT-4o reported 15 successes, 29 failures, and 1 warning, correctly identifying 11/15 successes and 28/29 failures.\\
\hline

\rowcolor{gray!15}
PS\#10 \cite{fathallah2025accessguru} &
Introduce AccessGuru to detect/fix accessibility violations and AccessCode dataset &
AccessGuru multi-agent framework vs Baseline LLM vs Human Professionals &
Real-world websites (10 representative sites) &
GPT-4, AccessGuru (GPT-4) &
GPT-4 &
Agent vs Baseline &
Detection Accuracy: AccessGuru (82\%) vs. Standard LLM (31\%). Correction Accuracy: AccessGuru (74\%) vs. Standard LLM (16\%). \\
\hline

PS\#11 \cite{doush2024evaluating} &
Evaluate accuracy of ChatGPT and Copilot in producing accessible web code &
Evaluating code generated by chatbots against "ideal" code metrics &
Web page features &
ChatGPT 3.5; Copilot &
ChatGPT 3.5 &
head-to-head &
ChatGPT 3.5: Valid Code 96.30\%; Correct Code 77.78\%; Screen Reader Compatible 55.56\%. Copilot: Valid Code 88.89\%; Correct Code 62.96\%; Screen Reader Compatible 48.15\%. \\
\hline

\rowcolor{gray!15}
PS\#12 \cite{cao2025scenegena11y} &
Propose SceneGenAlly to improve 3D scene accessibility  &
"Acc3D" agent vs No Agent baseline in a 3D environment &
3D web scenes &
GPT-4o &
GPT-4o &
baseline vs LLM &
The Acc3D agent enabled an 85\% success rate for accessibility tasks, compared to 0\% without the agent. \\
\hline

PS\#13 \cite{anderer2025making} &
Introduce LectureAssistant (AI chatbot) for blind/low-vision students in lectures &
User study with LectureAssistant prototype (thinking aloud, tasks) &
Education (Lecture video platform) &
Llama 3.1, MiniCPM-V &
N/A &
N/A &
No direct numeric comparison between Llama 3.1 and MiniCPM-V; evaluation focused on the tool's utility versus existing platforms. \\
\hline

\rowcolor{gray!15}
PS\#14 \cite{li2025videoa11y} &
Introduce VideoA11y (MLLMs) to generate video descriptions and VideoA11y-40K dataset &
Multi-study evaluation (Sighted and BLV users) of VideoA11y method &
YouTube and TikTok (video-sharing platforms) &
GPT-4V, GPT-4o, Video-LLaVA, LLaVA-OneVision, VILA &
GPT-4V &
 &
GPT-4V significantly outperformed Video-LLaVA on descriptiveness, accuracy, and clarity (p<0.05), with no significant difference for objectiveness (p>0.05). \newline
BLV users preferred VideoA11y descriptions 90\% of the time and rated them significantly higher in satisfaction (8.54 vs 5.43; p<0.001). \\
\hline

PS\#15 \cite{mo2025tablenarrator} &
Introduce TableNarrator using AI to generate alt text for image tables for BLV users &
Technical evaluation and User study of TableNarrator system &
Not reported &
GPT-4, GPT-4V , Claude 3 &
GPT-4 &
head-to-head &
TableNarrator (GPT-4) outperformed baselines (ROUGE-L = 0.360; selection = 66.86\%), exceeding Claude 3 (0.205; 21.14\%) and GPT-4V (0.240; 12.00\%), and achieved higher row/column accuracy (0.83\% vs. 0.05\% and 0.19\%). \\
\hline

\rowcolor{gray!15}
PS\#16 \cite{cheema2025describe} &
Explore user-driven AI audio descriptions allowing BLV users to control timing/detail &
User study evaluating user-driven AD levels (concise vs detailed) &
Online video (YouTube, TikTok, Instagram) &
GPT-4V &
N/A &
N/A &
GPT-4V (User-driven): Median efficiency=4, effectiveness=4, enjoyability=3; Concise AD activated more often (Mean=5.42) than detailed (Mean=3.58). \\
\hline

PS\#17 \cite{vera2025accessible} &
Workflow using LLMs to remediate inaccessible web content to "born-accessible"  &
Generation of remediated webpages &
Static university pages &
GPT-4o, Gemini &
Gemini &
head-to-head &
Gemini produced significantly fewer WAVE errors (p = 0.0435) and alerts (p = 0.0151) compared to GPT-4o, though Lighthouse scores showed no significant difference. \\
\hline

\rowcolor{gray!15}
PS\#18 \cite{pedemonte2025improving} &
AlternAtIve: Browser extension using LLMs to generate alt text for STEM images  &
Comparison of extension (Gemini) vs other tools &
Not reported &
Gemini 1.5 Pro , ChatGPT-4 &
Gemini 1.5 Pro &
tool vs tool &
AlternAtIve [H] (using Gemini 1.5 Pro) achieved the highest quality scores (Qua1 mean 0.91), significantly outperforming other treatments (p < 0.01 in most pairs). \\
\hline

PS\#19 \cite{andruccioli2025leveraging} &
Investigate LLMs' ability to detect accessibility issues in dynamic web content &
Analysis of handcrafted code and student projects &
Web application components &
GPT-4o mini, o3-mini, Gemini &
Gemini &
head-to-head &
Gemini had a much lower hallucination rate (20.1\%) compared to GPT-4o mini (69.9\%) in detecting violations. o3-mini achieved the highest self-verification accuracy (79\%), significantly outperforming GPT-4o mini (34\%) and Gemini (30\%). \\
\hline

\rowcolor{gray!15}
PS\#20 \cite{andruccioli2025tabular} &
To benchmark LLMs on detecting/remediating code accessibility issues &
Students generated components using LLMs &
Not reported &
GPT-4o mini, o3-mini, and Gemini 2.0 Flash &
N/A &
N/A &
The paper states that GPT-4o mini, o3-mini, and Gemini 2.0 Flash were evaluated in prior work using this dataset, but provides no metrics, numeric results, or comparative outcomes in this article. \\
\hline


PS\#21 \cite{lopez2025turning} &
Investigate if LLMs can automate manual WCAG success criteria evaluation &
LLM-powered scripts evaluated on WCAG SCs &
Not reported &
GPT-3.5, GPT-4, Claude 2, Google Bard &
N/A &
N/A &
The study reports that LLM scripts achieved 78\%/85\%/100\% success across the three WCAG SC, with 87.18\% overall (34/39). SC 1.1.1 (GPT-3.5): 78\% successful. SC 2.4.4 (Claude): 85\% successful. SC 3.1.2 (GPT-4): 100\% successful. \\
\hline

\rowcolor{gray!15}
PS\#22 \cite{afsal2025websumm} &
Chrome extension using local LLMs to summarize web content for the visually impaired &
Computational evaluation and Usability study &
BBC News articles across domains &
Llama 3.1, Phi 3, Gemma 2, and Mistral &
Gemma 2 &
head-to-head &
Computational evaluation: Gemma 2 led in overall score (0.585) and readability (FRE 52.53), while Mistral excelled in semantic similarity (0.8033).
User preference: Gemma 2 was most preferred (35.2\%).\\
\hline

PS\#23 \cite{michalichem2025gamified} &
Case study using LLMs with interface images to assist in heuristic usability evaluations for ID users &
Evaluation of "Add Task" and "Add Habit" screens &
Gamified web solution for task management and habit formation &
GPT-4o, DeepSeek V3, Claude 3.7 Sonnet &
Claude 3.7 Sonnet &
head-to-head &
Claude (3.7 Sonnet): Found most issues (16 in Exp 1, 17 in Exp 2); Highest unique issues (6 in Exp 1, 10 in Exp 2). ChatGPT (GPT-4o): Unique issues (5 in Exp 1, 6 in Exp 2). DeepSeek (V3): Unique issues (2 in Exp 1, 1 in Exp 2). Gemini (2.0 Flash): Fewest issues found (7 in Exp 1, 12 in Exp 2); Unique issues (2 in Exp 1, 1 in Exp 2). \\
\hline

\rowcolor{gray!15}
PS\#24 \cite{abu2025can} &
Benchmark generative-AI coding tools on producing accessible HTML/web &
Systematic testing with prompts and follow-ups &
Not reported &
ChatGPT 4o,Grok 3, Copilot Pro, Claude 3.7 Sonnet &
Claude 3.7 Sonnet &
head-to-head &
Claude 3.7 Sonnet and Copilot Pro tied for highest first-attempt success (72.7\%), but Claude was most efficient (fewest follow-up prompts) and had fewer severe violations than Grok 3. \\
\hline


PS\#25 \cite{afsal2024comparative} &
Compare accessibility and readability of ChatGPT vs. Google BARD chatbots &
Evaluation of interface accessibility and output readability &
Conversational AI chatbot interfaces &
GPT 3.5, BARD &
Google BARD &
head-to-head &
BARD passed AChecker but had 26 WAVE ARIA issues, while GPT failed AChecker but had 0 WAVE ARIA issues. Google BARD consistently outperformed GPT in readability, yielding higher scores across most measures. \\
\hline

\rowcolor{gray!15}
PS\#26 \cite{singh2024accessibility} &
Analyze education website accessibility for optimization &
Case study optimizing a sample page &
Education board websites  &
Claude &
N/A &
N/A &
Claude: Successfully improved accessibility by increasing color contrast (e.g., \#ccc to \#fff) and adding focus outlines. \\
\hline

PS\#27 \cite{silva2024page} &
Tool using BERT/ChatGPT to generate navigation aids for screen-reader users &
User study with simulated prototype and browser extension &
Text-only web pages from news domains &
BERT, GPT-3 &
N/A &
N/A &
BERT achieved the lowest error rate (8\%) on long documents, outperforming C99, while GPT-3 produced more coherent, descriptive headers than keyword-based baselines.\\
\hline

\rowcolor{gray!15}
PS\#28 \cite{delnevo2024interaction} &
Investigate ChatGPT for evaluating and correcting HTML accessibility issues &
Qualitative analysis of LLM responses to code snippets &
Not reported &
GPT3.5 &
N/A &
N/A &
GPT-3.5: Correctly identified well-structured forms and missing labels; flagged issues in tables but suggested semantics-altering fixes; correctly noted context for empty alt attributes. \\
\hline

PS\#29 \cite{huynh2024smartcaption} &
Extension using LLMs to generate context-aware image descriptions &
Evaluation of 10 images by participants &
News articles &
GPT-4o, AI-MCS &
SmartCaption AI (GPT-4o) &
Tool vs Tool &
Experiment A: SmartCaption AI outperformed baselines (8.3/10 vs. 3.6/10 and 1.7/10), which frequently hallucinated objects.
Experiment B: Initial latency was high (11.11s) but dropped to ~2.54s with caching. \\
\hline

\rowcolor{gray!15}
PS\#30 \cite{oswal2024examining} &
Evaluate the accessibility of three GenAI website builders and provide best practices &
Expert and automated testing of generated sites &
AI Website Builder Platforms / Artificial Design Intelligence tools. &
Dorik.com, Relume.io, and Wix.com &
Relume.io &
Generative AI Website Builders &
While Relume.io was the only tool rated "Partially accessible" in the screen reader walkthrough and produced fewer WAVE alerts than the others, all three platforms exhibited significant accessibility failures, such as high rates of unlabeled buttons and lack of WCAG 2.1 AA compliance. \\
\hline

PS\#31 \cite{acosta2024addressing} &
Evaluate web accessibility of 20 GenAI educational applications &
Automated (WAVE) and manual review of 20 tools &
20 generative artificial intelligence (AI) applications &
ChatGPT, Midjourney, Copilot, Suno 
&
Midjourney &
tool vs tool &
Midjourney: 1 WAVE error; 6 contrast errors. Suno: 1 WAVE error; 27 contrast errors. ChatGPT: 3 WAVE errors; 0 contrast errors; 1 alert. Copilot: 3 WAVE errors; 0 contrast errors. \\
\hline

\rowcolor{gray!15}
PS\#32 \cite{moon2024sagol} &
To generate alt text and search images for BLV users  &
 &
 &
MiniGPT-4, BLIP &
MiniGPT-4 &
head-to-head &
MiniGPT-4 produced longer alt text and higher top-10 accuracy (68.7\%) than BLIP (64.3\%) and object-recognition baselines (13.3–24.3\%). \\
\hline

PS\#33 \cite{othman2023fostering} &
Evaluate ChatGPT’s effectiveness in fixing WCAG 2.1  &
Using ChatGPT to fix 39 identified errors &
Qatar \& British Airways &
ChatGPT &
ChatGPT &
N/A &
ChatGPT successfully remediated 37 out of 39 identified errors across the two websites, achieving an accuracy rate of 94\%. \\
\hline

\end{tabular}%
}
\end{table*}

\section{Discussion}
\label{sec:Discussions}

Across the reviewed studies, LLMs are predominantly applied as task-oriented assistive components rather than as holistic accessibility solutions. Most approaches focus on supporting narrowly scoped activities—such as generating alternative text~\cite{li2025videoa11y,pedemonte2025improving,huynh2024smartcaption,moon2024sagol}, identifying accessibility violations~\cite{he2025enhancing,duarte2025expanding,paterno2025llm,andruccioli2025leveraging,lopez2025turning}, or suggesting localized fixes—embedded within existing development or evaluation workflows~\cite{gurita2025llm,fathallah2025accessguru,vera2025accessible,othman2023fostering}. This pattern suggests that current research conceptualizes LLMs as augmentation tools rather than as autonomous or end-to-end accessibility systems. While this reflects a pragmatic, risk-aware adoption strategy, it also limits exploration of how LLMs might support broader accessibility lifecycles, including continuous monitoring, iterative remediation, and user-facing accessibility assistance.

Simultaneously, the diversity of application forms—ranging from prompt-based usage to automated tools and browser extensions—indicates growing interest in integrating LLMs at different points of interaction. However, the limited number of standalone systems and benchmarks suggests that the field has not yet converged on shared infrastructures or standardized platforms for deploying and evaluating LLM-based accessibility support at scale.

\textbf{Alignment with Accessibility Standards and Disability Coverage.} The review shows strong reliance on WCAG as the primary framework for defining and operationalizing accessibility, with WCAG success criteria frequently used to guide task selection, evaluation, and reporting~\cite{he2025enhancing,paterno2025llm,fathallah2025accessguru,doush2024evaluating,vera2025accessible,andruccioli2025leveraging,lopez2025turning,abu2025can,oswal2024examining,acosta2024addressing,othman2023fostering}. In contrast, COGA guidelines and other disability-specific frameworks are referenced far less often~\cite{panchanadikar2025can}, resulting in uneven coverage across disability groups. Most studies prioritize accessibility needs related to visual impairments and screen reader compatibility, while cognitive, learning, and neurodiverse accessibility needs remain underrepresented.

This imbalance suggests that current LLM-based accessibility research largely mirrors long-standing trends in web accessibility, where visually observable or machine-detectable issues dominate. Although WCAG alignment provides consistency and comparability, limited engagement with COGA highlights an opportunity—and a need—to expand accessibility research beyond compliance-driven perspectives toward more inclusive and cognitively informed design considerations.

\textbf{Technical Design Choices.} From a technical standpoint, the literature is characterized by concentration rather than diversity. Most studies rely on a small set of commercial, general-purpose LLMs and relatively simple prompting strategies, with limited experimentation in model adaptation, architectural complexity, or long-running systems~\cite{panchanadikar2025can,he2025enhancing,aljedaani2025enhancing,duarte2025expanding,doush2024evaluating,abu2025can,afsal2024comparative,delnevo2024interaction,othman2023fostering}. Prompt engineering is frequently treated as the primary mechanism for improving accessibility outcomes, while architectural decisions—such as multi-agent coordination, feedback loops, or integration with accessibility tooling—are less commonly explored~\cite{duarte2025expanding,fathallah2025accessguru,vera2025accessible,andruccioli2025leveraging,lopez2025turning,huynh2024smartcaption}.

This focus reflects both the accessibility of current LLM APIs and the exploratory nature of the field. However, it also constrains the types of accessibility problems that can be effectively addressed. More complex accessibility needs—such as context-sensitive guidance, personalization, or sustained interaction—may require architectural designs that go beyond single-prompt, single-response paradigms. The current emphasis on simplicity may therefore limit progress toward robust, real-world accessibility support.

\textbf{Evaluation Practices and Evidence Quality.} Evaluation practices across the literature are methodologically diverse but uneven. While many studies employ empirical or experimental evaluations, these are often conducted in controlled settings using automated tools, expert review, or small-scale participant samples (Table \ref{tab:participants_verbatim}). Direct involvement of people with disabilities remains limited, and when present, participant groups are often small or heterogeneous. As a result, much of the reported evidence emphasizes technical feasibility and comparative performance rather than real-world usability or lived experience~\cite{duarte2025expanding,fathallah2025accessguru,li2025videoa11y,abu2025can,afsal2024comparative,huynh2024smartcaption}.

For a domain where user impact is central, this raises concerns about the ecological validity of current findings. The limited use of participatory or longitudinal evaluation approaches suggests that accessibility research involving LLMs has yet to fully embrace user-centered and inclusive evaluation practices. Strengthening this aspect of research will be critical to ensuring that LLM-based accessibility tools genuinely benefit the users they aim to support.

\textbf{Prioritized Accessibility Issues and Emerging Challenges.} The synthesis reveals that LLM-based accessibility research concentrates heavily on perceivability and semantic clarity, particularly issues involving missing alternative text, insufficient color contrast, labeling, and structural markup~\cite{panchanadikar2025can,duarte2025expanding,fathallah2025accessguru,li2025videoa11y,pedemonte2025improving,lopez2025turning,abu2025can,huynh2024smartcaption,moon2024sagol}. These issues are well-suited to automated detection and natural language generation, which likely explains their prominence~\cite{he2025enhancing,paterno2025llm,lopez2025turning}. In contrast, accessibility challenges that involve temporal behavior, interaction complexity, or cognitive load—such as time-based media accessibility, navigation consistency, reading level, and content comprehensibility—are addressed far less frequently~\cite{panchanadikar2025can,anderer2025making,li2025videoa11y,andruccioli2025leveraging,afsal2025websumm,michalichem2025gamified,afsal2024comparative}.

Importantly, the review also identifies emergent, LLM-specific accessibility issues, such as hallucinated descriptions, misleading compliance claims, and incorrect ARIA usage. These problems are not always captured by existing WCAG success criteria, indicating that generative systems introduce new classes of accessibility risks. Addressing these risks may require extending current standards, developing new evaluation heuristics, or rethinking how accessibility compliance is assessed in the presence of AI-generated content.

\section{Research Implications}
\label{sec:ResearchImplications}
This section discusses the implications of our findings for both research and practice in the context of LLM-supported web accessibility. We highlight directions for future research and practical considerations for responsibly integrating LLMs into accessibility workflows.

\subsection{Implications for Research}
Future research should broaden the scope of accessibility issues addressed by LLM-based approaches, particularly by engaging more deeply with cognitive accessibility, neurodiversity, and time-based media accessibility. Greater attention to COGA guidelines and other disability-specific frameworks would help ensure more inclusive coverage across user needs.

Researchers should also explore architectural diversity beyond single-prompt or single-model designs, including multi-stage pipelines, adaptive systems, and hybrid approaches that combine LLMs with rule-based accessibility tools. Such exploration is essential to understanding how LLMs can support sustained, context-aware accessibility workflows. Evaluation practices must move beyond feasibility demonstrations toward rigorous, user-centered, and participatory methods. Involving people with disabilities as primary stakeholders throughout design and evaluation is critical to producing valid and actionable insights.

\subsection{Implications for Practice}
For practitioners, the findings suggest that LLMs can be valuable in supporting specific accessibility tasks—such as generating alternative text or identifying common accessibility violations—but should be used with caution. LLM outputs should be treated as assistive suggestions rather than authoritative solutions, particularly given the risk of hallucinated or incomplete guidance.

Practitioners should also be aware that current LLM-based tools tend to emphasize compliance with easily detectable WCAG criteria, while offering limited support for cognitive and experiential aspects of accessibility. Integrating LLMs into accessibility workflows, therefore, requires complementary practices, including expert review, user testing, and adherence to established accessibility standards. In general, both researchers and practitioners should approach LLM-based accessibility support as a collaborative, human-in-the-loop process, ensuring that automation enhances rather than undermines inclusive web experiences.

\section{Threats to Validity}
\label{sec:ThreatsToValidity}
As with any systematic literature review, this study is subject to several threats to validity. We outline these threats below, along with the steps taken to mitigate them.

\textbf{Selection Validity.} Relevant studies may have been missed due to limitations in the search strategy, choice of digital libraries, or keyword formulation. To mitigate this risk, we searched multiple widely used digital libraries, iteratively refined the search strings through pilot searches, and applied backward and forward snowballing. Nonetheless, some relevant studies—particularly those using non-standard terminology or published in non-indexed venues may not have been captured.

\textbf{Construct Validity.} Threats to construct validity arise from variations in how primary studies define and operationalize concepts such as web accessibility, LLM-based approaches, and accessibility issues. To mitigate these threats, we relied on the explicit definitions and descriptions provided by the authors of the primary studies and systematically mapped reported issues to established frameworks such as WCAG and COGA, where applicable, to ensure consistent categorization.

\textbf{Internal Validity.} The manual filtering, data extraction, and synthesis processes may introduce researcher bias or inconsistent interpretation. To mitigate this threat, two authors independently reviewed the extracted data, resolved disagreements through discussion, and documented decisions in a shared spreadsheet. The authors’ prior research experience in web accessibility further supported consistency during the review process.

\section{Conclusion } 
\label{sec:Conclusion}

This study presents a systematic literature review of 38 peer-reviewed studies investigating the use of Large Language Models to support web accessibility. Through our analysis, we identify the accessibility tasks addressed by these studies, the LLM-based approaches and system designs employed, and the accessibility issues and standards considered. Our findings show that most studies apply LLMs to support a limited set of accessibility tasks—primarily content generation, issue detection, and semantic remediation—with WCAG serving as the dominant reference framework and limited incorporation of cognitive accessibility guidelines such as COGA.

Further analysis highlights common patterns across the reviewed studies, including a strong reliance on general-purpose LLMs, prompt-based interaction strategies, and relatively simple system architectures. Evaluation practices vary substantially, with many studies relying on automated assessments or expert inspection, while fewer involve users with disabilities. We envision that our findings serve as a consolidated reference for researchers and practitioners seeking to understand the current landscape of LLM-supported web accessibility. Future work in this area includes broader coverage of accessibility needs, particularly cognitive accessibility; more rigorous user-centered evaluation; and the development of benchmarks and evaluation frameworks tailored to the accessibility challenges posed by generative AI systems.

\balance

\bibliographystyle{ACM-Reference-Format}
\bibliography{reference}

\end{document}